\documentstyle[12pt,psfig]{article}
\title{\bf Chemical Abundances of Bright Giants in the Mildly Metal-Poor 
Globular Cluster M4}
\author{Inese I. Ivans\thanks{iivans@astro.as.utexas.edu}
\vspace{1cm}\\
\normalsize Department of Astronomy and McDonald Observatory, University of Texas at Austin, USA}
\date{\mbox{}}
\begin{document}
\maketitle
\pagestyle{empty}
%
% WE REDEFINE THE plain LaTeX PAGESTYLE !!! 
% THIS PAGESTYLE WILL BE USED FOR THE FIRST PAGE ONLY !
%
\def\bull{\vrule height .9ex width .8ex depth -.1ex}
\makeatletter
\def\ps@plain{\let\@mkboth\gobbletwo
\def\@oddhead{}\def\@oddfoot{\hfil\tiny\bull\quad
``The Galactic Halo: From Globular Clusters to Field Stars'';
35$^{\mbox{\rm th}}$ Li\`ege\ Int.\ Astroph.\ Coll., 1999\quad\bull}%
\def\@evenhead{}\let\@evenfoot\@oddfoot}
\makeatother
%
% AND DEFINE OUR MACROS FOR THE REFERENCE LIST
% I.E \beginrefer \refer and \endrefer
%
\def\beginrefer{\section*{References}%
\begin{quotation}\mbox{}\par}
\def\refer#1\par{{\setlength{\parindent}{-\leftmargin}\indent#1\par}}
\def\endrefer{\end{quotation}}
%
% BEGIN THE ABSTRACT CHAPTER WITH \noindent\small, ENCLOSE IT IN A GROUP
% AND BOLDFACE THE TITLE.
%
{\noindent\small{\bf Abstract:} 

We present a chemical composition analysis of three dozen giant stars in 
the nearby ``CN-bimodal'' mildly metal-poor ($<$[Fe/H]$>$ = -1.18) globular 
cluster M4.  The analysis combined traditional spectroscopic abundance 
methods with modifications to the line-depth ratio technique pioneered by 
Gray (1994).  Silicon and aluminum are found to be primordially overabundant 
by factors exceeding the mild overabundances usually seen in $\alpha$- and 
light odd elements among halo field and globular cluster giants of 
comparable metallicity.  In addition, barium is found to be overabundant by 
a factor of about four.  Superimposed on the primordial abundance 
distribution in M4, there is evidence for the existence of proton-capture 
synthesis of C, O, Ne, and Mg.}
%
% NOW COMES THE MAIN BODY OF THE ARTICLE
%
\section{Introduction}

As isolated laboratories of stellar evolution, individual globular clusters 
were once considered to be simple systems, having formed coevally, out of 
the same material, and exhibiting cluster-to-cluster differences due to 
only metallicity and age effects.  In reality, clusters of similar age and 
metallicity exhibit differences in their colour-magnitude diagrams and many 
of the elemental abundance patterns deviate from the predictions of stellar 
evolution theory.  Many low-metallicity globular clusters exhibit large 
star-to-star variations of C, N, O, Na, Mg, and Al abundances.  These 
elements are those that are sensitive to proton-capture nucleosynthesis.  

In clusters where giant star samples have been sufficiently large, the 
abundances of O and Na are anticorrelated, as are those of O and Al (as
well as sometimes Mg and Al).  Previous clusters studied by the Lick-Texas 
group (including M3, M5, M10, M13, M15, M71, M92, and NGC7006) span a 
range in metallicities, from --0.8 $\le$ [Fe/H] $\le$ --2.24.  In the 
higher-metallicity clusters, the abundance swings are muted.  In all of 
the clusters, the abundance swings are observed to be a function of giant 
branch position.  This relationship is consistent with material having 
undergone proton-capture nucleosynthesis (via the CN-, ON-, NeNa-, and/or 
MgAl-cycles) and brought to the surface by a deep-mixing mechanism.  Deep
mixing, according to theory (Sweigart \& Mengel 1979) should become less
efficient and possibly cut-off as metallicity increases.  The metallicity 
of M4 places it among clusters in which the O versus Na and Mg versus Al 
anticorrelations might be expected to be largely diminished.    

There are also some clusters (including M5, NGC3201, NGC6752, 47~Tuc, and 
$\omega$~Cen) for which distinctly bimodal distributions of cyanogen 
strengths at nearly all giant branch positions have been uncovered.  These 
clusters apparantly have had different nucleosynthesis histories.  Among 
these CN-bimodal clusters is M4 (Norris 1981; Smith \& Norris 1993), the 
nearest, brightest, and one of the most accessible targets to study the 
CN-bimodal phenomena.  For the purposes of this colloquium, only certain 
aspects of our M4 work will be highlighted.  Details of the full analysis 
are presented in Ivans {\it et al} (1999).

\section{Getting the Red Out}

While M4 may be the closest globular cluster, it also suffers from 
interstellar extinction that is large and variable across the cluster face.  
The line-of-sight to the cluster passes through the outer parts of the 
Scorpius-Ophiuchus dark cloud complex.  A reddening gradient exists across 
the face of the cluster (Cudworth \& Rees 1990; Liu \& Janes 1990; Minniti 
{\it et al} 1992).  And, the dust extinction probably varies on small 
spatial scales as well.  This is suggested  by the colour-magnitude diagram 
of M4, where the subgiant and giant branches are broader than expected, 
given the errors in the photometry (see figure 1) as well as by observations 
of the $\lambda7699\AA$ K~{\sc i} interstellar line towards individual M4 
stars (Lyons {\it et al} 1995). Figure 1 also shows that the reddening 
cannot reliably be estimated for individual M4 stars to the level needed to 
map broad-band photometric indices onto stellar parameters.  Instead, 
another reliable temperature estimation method is required.

%\vspace{5in}
%\special{psfile=1hr.eps angle=0 hoffset=-40 voffset=-10
%        vscale=100 hscale=100}
\begin{figure}[bhpt]
\centerline{\psfig{file=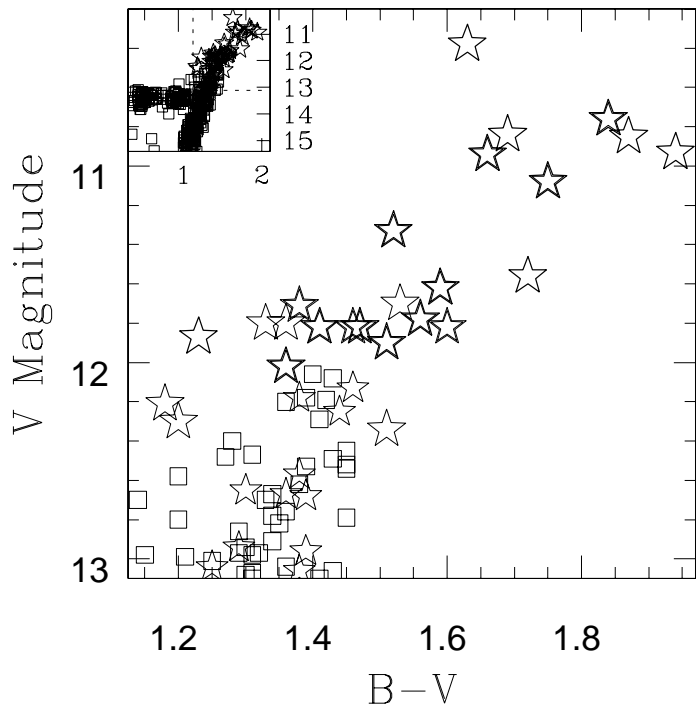,clip=}}
\caption{A colour-magnitude diagram of M4, with photometry from Cudworth \& 
Rees (1990), showing the positions of our program stars ($\star$) on the 
giant branch.  The inset diagram shows the program stars plotted in relation 
to all Cudworth \& Rees M4 stars of magnitude $\leq$~15.5 (adapted from Fig.
1 in Ivans {\it et al} 1999).}
\end{figure}

We combined traditional spectroscopic abundance methods with modifications 
to the line-depth ratio technique pioneered by Gray (1994) to determine the 
atmospheric parameters of our stars.  The ``Gray'' method relies on ratios 
of the measured central depths of lines having very different functional 
dependences on photometric indices and/or Teff to derive accurate relative 
temperature rankings (eg. vanadium versus neutral or ionized iron).  Gray's 
work was done on Pop.~I main sequence stars and has since been expanded by 
Hiltgen (1996) for applications to subgiants of a range of disk 
metallicities. Happily, many of Gray's line depth ratios are also sensitive 
Teff indicators for lower metallicity very cool RGB stars.  The line depth 
ratios vary more than one dex in spectra of giants of moderately metal-poor 
clusters, and thus can indicate very small Teff changes.  However, these 
relationships begin to approach unity among the coolest stars.  While a
tremendously useful tool, the ``Gray'' method cannot be applied to all
stars of all clusters: these ratios probably will be less useful as 
temperature indicators for the coolest stars of appreciably more metal-rich 
globular clusters (where the lower temperatures and higher metallicities 
conspire to saturate and blend virtually all of the ``Gray'' spectral 
features).  However, the method was successfully employed in our work on
M4 and is currently being applied to other clusters in the process of 
analysis.

Our initial Teff calibration of the M4 line depth ratios was set through 
a similar analysis of RGB stars of M5 (a cluster of very similar 
metallicity to M4 but suffers little from interstellar dust extinction).
We discuss the details of the correlations and transformations in Ivans 
{\it et al} (1999).  While we used the line-ratio method to rank the 
stars, final temperatures were determined from full spectral analyses.  
Our results for individual stellar parameters compare well with M4 stars in 
the literature.  Taking advantage of the non-photometric means by which we 
obtained our temperatures, we then derived an average $E(B-V)$ reddening of 
0.33 +/- 0.01 (which is significantly lower than that estimated by using 
the dust maps made by Schlegel {\it et al} 1998 but is in good agreement 
with the M4 RR Lyrae studies by Caputo {\it et al} 1985).    Finally, as 
a confirmation of the method, we derived individual stellar extinctions 
that not only correlate extremely well with IRAS 100 micron fluxes but 
also with $E(B-V)$ estimates derived independently in interstellar 
absorption studies of potassium by Lyons {\it et al} (1995).

\section{Abundance Results}

We performed line-by-line abundance analyses to determine the final model 
atmosphere specifications.  Our final models satisfied the following 
contraints: consistent abundances from lines of neutral and ionized Fe and 
Ti; reasonable predictions for colour-magnitude diagram positions from the 
derived gravities; no obvious trends of neutral Fe line abundances with 
EWs; and no obvious trends of neutral Fe line abundances with corresponding 
excitation potentials.  Finally, there is no astrophysical reason for 
Fe-peak abundances to vary significantly from star to star along the M4 
giant branch; V, Ti, Fe, and Ni showed no significant drifts along the RGB.

We present the abundance analyses in the ``boxplot'' shown in figure 2.
The boxplot illustrates the median, data spread, skew and distribution of 
the range of values we derived for each of the elements from our program 
stars, along with possible outliers.  We determined a metallicity of
$<$[Fe/H]$>$~= --1.18 ($\sigma$~=~0.02) and found a large abundance 
ratio range for proton-capture elements such as oxygen, sodium and aluminum.
However, the star-to-star variations are small for the heavier elements.  
Our M4 abundances generally agree well with those of past M4 investigators. 
The abundances of Ca, Sc, Ti, V, and Ni are also in accord with those of M5 
and the halo field.  However, the M4 abundances of Ba and La are both 
overabundant with respect to comparison samples.  Yet, the overabundance of 
Ba in M4 stars has been observed in independent studies by both Brown \& 
Wallerstein (1992) as well as by Lambert {\it et al} (1992).  We also derived 
high silicon and aluminum abundances, in agreement with previous studies of 
M4 but significantly higher than the abundances found in either M5 or the 
field.  We explore these issues in the following sub-sections.

%\vspace{4.3in}
%\special{psfile=1bp.eps angle=0 hoffset=0 voffset=0
%        vscale=1 hscale=1}
\begin{figure}[bhpt]
\centerline{\psfig{file=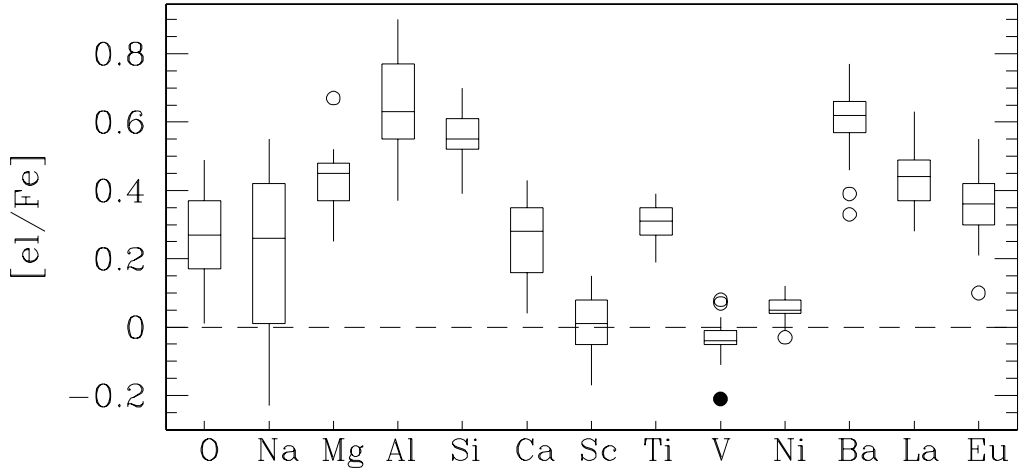,clip=}}
\caption{A ``boxplot'' of the M4 giant star element abundances.  For each
element, a boxed horizontal line indicates the median value and the 
interquartile range (the middle 50\% 
of the data).  The vertical tails extending from the boxes indicate the 
total range of abundances determined for each element, excluding outliers.  
Mild outliers (those between 1.5$\times$ and 3$\times$ the interquartile 
range) are denoted by hollow circles (o) and severe outliers (those greater 
than 3$\times$ the interquartile range) by filled circles ($\bullet$).  The 
dashed line at [el/Fe] represents the solar value for a particular 
elemental abundance ratio (taken from Fig. 7 in Ivans {\it et al} 1999).}
\end{figure}

\subsection{Proton-Capture Nucleosynthesis}

Although several M4 giants exhibit oxygen deficiencies, most M4 giants show 
little evidence for the severe oxygen depletions observed in M13 (Kraft 
{\it et al} 1997) and M15 (Sneden {\it et al} 1997).  Low oxygen 
abundances are accompanied by low carbon and elevated nitrogen.  In 
addition, the sum of C+N+O is essentially constant, as expected if all 
stars draw on the same primordial material.  We find that the behaviour of 
O is anti-correlated with that of Na, and, to a lesser degree, with that 
of Al.  These findings are compatible with a proton-capture scenario in 
which Na and Al are enhanced at the expense of Ne and Mg, respectively 
(Langer {\it et al} 1997, Cavallo {\it et al} 1998).  And, we find that 
the CN-strong stars are those that are more highly processed via 
proton-capture nucleosynthesis: the CN-strong group has a mean Na abundance 
that is a factor of two larger than the CN-weak group, and our CN-strong 
group also has higher Al abundances but the CN-strong/CN-weak difference is 
much less pronounced.  As shown by Langer \& Hoffman (1995), very modest 
hydrogen depletion of the envelope material can lead to an enhancement of 
Al by +0.4 dex when Na is enhanced by +0.7 dex, exactly as observed in our 
M4 sample. In this picture, unlike that found in M13 (Shetrone 1996), the 
enhancement of Al comes about entirely by destruction of $^{25}$Mg and 
$^{26}$Mg: $^{24}$Mg remains untouched. We attempted to derive the Mg 
isotope ratios in our spectra and, while there is a hint that the isotopic 
ratios may not be the same as those found in halo field stars, much higher 
resolution data is required before one can make statements regarding any 
differences with certainty.

\subsection{$\alpha$-element Enhancements}
Both the magnesium abundances and silicon abundances in M4 exceed those in
M5 by a factor of two.  However, Ca and Ti abundances in the two clusters are
essentially the same and have the usual modest overabundances with respect
to the scaled solar ratio.  The $\alpha$- element ratios in M4 mimic those
found in the very metal-poor cluster M15.  M15, like M4, also exhibits a
high aluminum abundance (that is, a high ``floor'' of aluminum, on top of
which is the proton-capture nucleosynthesic contribution described in the
previous subsection).  Substructure in $\alpha$- and light odd elements
are also found among relatively metal-rich disk dwarfs (eg. Edvardsson 
{\it et al} 1993) and galactic nuclear bulge giants (McWilliam \& Rich 
1994).  While the abundance pairings and trends are not matched between 
the cluster and disk/bulge populations, it is clear that the differences 
must arise from some property of the primordial nucleosynthetic sites.  

\subsection{The Abundances of Ba, La, Eu and $\omega$ Cen}

The [Ba/Eu] ratio is often used as a measure of $s$- to $r$-process 
nucleosynthesis in the primordial material of a cluster.  Typically,
clusters show --0.6 $<$ [Ba/Eu] $<$ --0.2.  In M4, [Ba/Eu] is  0.25~dex 
higher than the total solar-system {\it r~+~s} and more than four times 
higher than that of the ``normal" cluster M5.  However, the high [Ba/Eu] 
in M4 is not because Eu is low (as is the case in very metal-poor M15), 
rather, the [Eu/Fe] we find for M4 is not very different from that of M5.  
The high [Ba/Eu] is due to a high [Ba/Fe].  And, the high abundance of Ba 
is supported by a high abundance of La.  We performed numerical experiments 
by combining the results of our derived Ba, La, and Eu abundances and found 
that M4 has a larger $s$:$r$-process contribution than in the sun; the Ba
abundance in M4 cannot be attributed to the $r$-process.  We find no 
dependence of the Ba or La abundance on evolutionary state in M4
$\Rightarrow$ these 
excesses cannot result from neutron captures on Fe-peak elements during a 
He shell flash episode on the AGB of the stars we observed.  It {\it must} 
be a signature of $s$-process enrichment of the primordial material out of 
which the low-mass M4 stars we observed were formed.  This excess of the 
$s$-process elements is evidence that the period of star formation and 
mass-loss that preceded the formation of the observed stars in M4 was long 
enough for stars of 3--10 solar masses to evolve into AGB stars and 
contribute their ejecta into the ISM of the cluster.   $s$-process 
contributions such as those we found in M4 are very well evidenced in the 
globular cluster $\omega$ Cen (Vanture {\it et al} 1994).  Interestingly 
enough, there exists in $\omega$~Cen a subset of stars which possess 
nearly identical overabundance characteristics in [Ba/Fe], [Al/Fe], 
[Si/Fe], and [La/Fe] as those found in our M4 stars.  However, the 
multi-metallicity cluster $\omega$ Cen also possesses a more complicated 
nucleosynthetic history than M4.  The important point here is that the 
high Ba and La properties of M4 stars is surely a primordial, not an 
evolutionary, effect.

\section{Conclusions and Future Work}
Evidence for proton-capture nucleosynthesis in M4 was expected, and was 
found to be in good agreement with both observations and theory.  However, 
the overabundances of Ba, La, Si, and Al were not expected and these are 
still a puzzle.  While there are similarities in metallicity, and 
evolutionary age as observed in the colour-magnitude diagrams of M4 and 
M5, what nucleosynthetic histories can explain such large differences in 
the elemental abundances?  The Mg isotope ratio in M4 may also be found to 
be different from that of the field stars, yet another difference that the 
environment may have imposed on the nucleosynthetic history.  Many of the 
colloquium talks have emphasized the need to attack both more outer halo 
clusters as well as the disk clusters and do analyses similar to those that 
have been done for the closest halo clusters. With the successful 
application of the line-ratio techniques, more detailed abundance analyses 
will be {\it able} to be done.
%
% USE A SECTION WITHOUT NUMBER FOR THE ACKNOWLEDGEMENTS
%
\section*{Acknowledgements}
I am indebted to Jerry Lodriguss for generously sharing the electronic 
files containing his excellent deep sky images of the M4 Sco-Oph region.  
I also very gratefully acknowledge the contributions made by my collaborators
to this work, without whom the successful completion of this project 
would not have been possible.
%
% BEGIN THE REFERENCE LIST WITH \beginrefer
% USE \refer BEFORE THE REFERENCES AND BEGIN A NEW PARAGRAPH AFTER THE 
% REFERENCE !
% DO NOT FORGET TO END THE LIST WITH \endrefer
%
 
\beginrefer

\refer Brown, J. A. \& Wallerstein, G., 1992 AJ 104, 1818.

\refer Caputo, F., Castellani, V., \& Quarta, M. L., 1985, A\&A 143, 8.

\refer Cavallo, R. M, Sweigart, A. V., \& Bell, R. A., 1998, ApJ 492, 575.

\refer Cudworth, K. M. \& Rees, R. F., 1990, AJ 99, 1491.

\refer Edvardsson, B., Andersen, J., Gustafsson, B., Lambert, D. L., 
	Nissen, P. E., \& Tomkin, J., 1993, A\&A 275, 101.

\refer Gray, D. F.,1994, PASP 106, 1248.

\refer Gustafsson, B., Bell, R. A., Ericksson, K., \& Nordlund, A.,  1975, 
	A\&A 42, 407.

\refer Hiltgen, D.. 1996, Ph.D. Thesis, Univerity of Texas, Austin.

\refer Ivans, I. I., Sneden, C., Kraft, R. P., Suntzeff, N. B., Smith, V. V.,
       Langer, G. E., Fulbright, J. P., 1999, AJ 118, 1273.

\refer Kraft, R. P., Sneden, C., Smith, G. H., Shetrone, M. D., Langer, G. 
	E., \& Pilachowski, C. A., 1997, AJ 113, 279.

\refer Lambert, D. L., McWilliam, A. \& Smith, V. V., 1992, ApJ 386, 685.

\refer Langer, G. E. \& Hoffman, R., 1995, PASP 107, 1177.

\refer Langer, G. E., Hoffman, R., \& Zaidins, C. S., 1997 PASP 109, 244.

\refer Liu, T. \& Janes, K. A., 1990, ApJ 360, 561.

\refer Lyons, M. A., Bates, B., Kemp, S. N., \& Davies, R. D., 1995 MNRAS 
	277, 113.

\refer McWilliam, A. \& Rich, R. M., 1994  A\&AS 91, 749.

\refer Minniti, D., Coyne, G. V., \& Claria, J. J., 1992, AJ 103, 871.

\refer Norris, J., 1981, ApJ 248, 177.

\refer Schlegel, D. J., Finkbeiner, D. P., \& Davis, M., 1998, ApJ 500, 525.

\refer Shetrone, M. D., 1996, AJ 112, 2639.

\refer Smith, G. H. \& Norris, J. E., 199,3 AJ 105, 173.

\refer Sneden, C., 1973, ApJ 184, 839.

\refer Sneden, C., Kraft, R. P., Shetrone, M. D., Smith, G. H., Langer, 
	G. E., \& Prosser, C. F. 1997 AJ 114, 1964.

\refer Sweigart, A. V. \& Mengel, J. G., 1979, ApJ 229, 624.

\refer Vanture, A.D., Wallerstein, G. \& Brown, J.A., 1994, PASP 106, 835.

\endrefer           
\end{document}